\documentclass[prb,letterpaper,aps,floatfix,superscriptaddress,twocolumn]{revtex4-1}
\usepackage{graphicx}
\usepackage{amsmath}
\usepackage{subfigure}
\newcommand{\beq}{\begin{equation}}
\newcommand{\eeq}{\end{equation}}
\newcommand{\bk}{{{\bf{k}}}}

\newcommand{\bA}{{\bf{A}}}

\newcommand{\bsigma}{{\boldsymbol \sigma}}
\newcommand{\beqa}{\begin{eqnarray}}
\newcommand{\eeqa}{\end{eqnarray}}

\begin{document}
\title{Reply to Comment by Vanderbilt, Souza and Haldane: arXiv:1312.4200}
\author{Y. Chen}
\affiliation{Department of Physics and Astronomy, University of Waterloo, Waterloo, Ontario 
N2L 3G1, Canada}
\author{D.L. Bergman}
\affiliation{Department of Physics, California Institute of Technology, 1200 E. California Blvd, MC114-36, Pasadena, California 91125, USA}
\author{A.A. Burkov}
\affiliation{Department of Physics and Astronomy, University of Waterloo, Waterloo, Ontario 
N2L 3G1, Canada} 
\date{\today}
\begin{abstract}
We reassert our statement that the non-quantized Weyl node contribution to the anomalous Hall conductivity is not a Fermi surface property, 
contrary to the claim by Vanderbilt, Souza and Haldane. 
\end{abstract}
\maketitle
We have recently pointed out~\cite{Burkov13} that the common statement in the literature that ``intrinsic anomalous Hall conductivity is entirely a Fermi surface 
property modulo a quantized contribution from filled bands" is incorrect, since it ignores non-quantized Weyl node contribution from filled bands. 
Vanderbilt, Souza and Haldane (VSH)~\cite{VSH13} assert that Weyl node contribution is included in the Fermi surface contribution and our claim stems from a mathematical 
error in the identification of the Fermi surface contribution. Here we show that no mathematical errors are involved in our claim, and our disagreement is based 
on physics, namely the fact that Weyl node contribution is associated with Fermi arc edge states and not with bulk states on the Fermi surface.  This fact is ignored
by VSH. 

To understand the nature of our disagreement with VSH, let us consider the following toy model Hamiltonian of a metallic ferromagnet with two Weyl nodes, 
i.e. the smallest possible number:
\beq
\label{eq:1}
H_t(\bk) = k_x \sigma_x + k_y \sigma_y + m_t(k_z) \sigma_z, 
\eeq
where $\bsigma$ are Pauli matrices, describing the two bands that touch at the Weyl nodes and will use $\hbar = 1$ units everywhere. 
The index $t=1,2$ labels two distinct Dirac fermions. The ``mass term" $m_1(k_z)$ is such that 
it changes sign from positive to negative at $k_z = \pm k_0$, which are the Weyl node locations, while mass term $m_2(k_z)$ is always positive. 
This is the simplest possible model of the electronic structure of a three-dimensional ferromagnet with Weyl nodes. 
The eigenvalues and corresponding eigenfunctions of $H_t$ are given by
\beqa
\label{eq:2}
\epsilon_{st \bk}&=&s \epsilon_{t \bk} = s \sqrt{k_x^2 + k_y^2 + m^2_t(k_z)}, \nonumber \\
|u_{s t\bk} \rangle&=&\frac{1}{\sqrt{2}} \left[\sqrt{1 + s \frac{m_t(k_z)}{\epsilon_{t \bk}}}, s \hat k_+ \sqrt{1 - s \frac{m_t(k_z)}{\epsilon_{t \bk}}} \right], 
\eeqa 
where $s = \pm$.
The components of the Berry connection field and the Berry curvature are easily calculated and given by
\beqa
\label{eq:3}
A^{s t}_x(\bk)&=&-i \langle u_{s t \bk}| \partial_{k_x} |u_{s t \bk} \rangle = - \frac{k_y}{2 \epsilon_{t \bk} [\epsilon_{t \bk} + s m_t (k_z)]}, \nonumber \\
A^{s t}_y(\bk)&=&-i \langle u_{s t \bk}| \partial_{k_y} |u_{s t \bk} \rangle = \frac{k_x}{2 \epsilon_{t \bk} [\epsilon_{t \bk} + s m_t (k_z)]}.
\eeqa
and 
\beq
\label{eq:4}
\Omega^{s t}_z(\bk) = \partial_{k_x} A^{s t}_y(\bk) - \partial_{k_y}A^{s t}_x(\bk) = s \frac{m_t(k_z)}{2 \epsilon_{t \bk}^3};
\eeq
Let us take the Fermi energy to be somewhere in the upper $s = +$ bands, i.e. $\epsilon_F >0$. 
The anomalous Hall conductivity can now be calculated as integral of the Berry curvature over all occupied states. 
Let us separately consider the contributions of the fully occupied $s = -$ bands and the partially occupied $s = +$ bands. 
We have
\beqa
\label{eq:5}
\sigma^-_{xy}&=&e^2 \sum_t \int \frac{d^3 k}{(2 \pi)^3} \Omega^{- t}_z(\bk) \nonumber \\
&=&- \frac{e^2}{8 \pi^2} \sum_t \int d k_z \, \textrm{sign}[m_t(k_z)] = - \frac{e^2 k_0}{2 \pi^2}, 
\eeqa
and
\beqa
\label{eq:6}
\sigma^+_{xy}&=&e^2 \sum_t \int \frac{d^3 k}{(2 \pi)^3} \Omega^{+ t}_z(\bk) n_F(\epsilon_{t \bk}) \nonumber \\
&=&\frac{e^2}{8 \pi^2} \sum_t \int d k_z \left[\textrm{sign}[m_t(k_z)] - m_t(k_z)/\epsilon_F \right] \nonumber \\
&\times&\Theta(\epsilon_F - |m_t(k_z)|).
\eeqa
where $\Theta(x)$ is the Heaviside step function. 
The $\sigma^-_{xy}$ contribution, coming from completely filled bands, is precisely what we call the Weyl node contribution. 
Note that $\sigma^+_{xy}$ is a continuous function of the Fermi energy and cancels the Weyl node contribution 
$\sigma^-_{xy}$ only in the limit $\epsilon_F \rightarrow \infty$, i.e. when the upper bands are filled. 

Let us now focus on the contribution of the unfilled bands, $\sigma^+_{xy}$. 
We first note that the circulation of the Berry connection field along a section of the Fermi surface (i.e. the total Berry phase accumulated along this section), 
corresponding to a specific value of $k_z$, is given by
\beqa
\label{eq:7}
&&\oint d \bk \cdot \bA^{+ t}(\bk) = \pi \left[1 - m_t(k_z)/\epsilon_F\right] \nonumber \\
&=&\pi \left[\textrm{sign}[m_t(k_z)] - m_t(k_z)/\epsilon_F\right] + \pi[1 - \textrm{sign}[m_t(k_z)]]. \nonumber \\
\eeqa
Note that the last term in Eq.~\eqref{eq:7} is either equal to $0$ or $2 \pi$, which is a reflection of a $2 \pi$ ambiguity 
of the definition of Berry phase.  

The essence of VSH's argument is that one can use the above $2 \pi$ ambiguity of the Berry phase 
to include the Weyl node contribution $\sigma^-_{xy}$ into the Fermi surface contribution as
\beq
\label{eq:8}
\sigma_{xy}^{FS} = \sigma_{xy}^+ - \frac{e^2}{8 \pi^2} \sum_t \int d k_z \, \textrm{sign}[m_t(k_z)] .
\eeq
By construction, this cancels out $\sigma_{xy}^-$ and everything is now a Fermi surface contribution. 
While this is formally correct, we see absolutely no physical reason for doing this and instead identify 
$\sigma_{xy}^{FS} = \sigma_{xy}^+$,~\cite{Burkov13}  which is equivalent to ignoring the second term in Eq.~\eqref{eq:7}. 
This is just as correct formally as what VSH propose, but in addition is also correct physically. 
The reason is that the contribution of filled bands $\sigma_{xy}^-$ can not be associated with Fermi surface and instead comes from 
chiral Fermi arc edge states, as is now well-understood in the context of extensive work on Weyl semimetals.~\cite{Wan11,Burkov11}
We thus conclude that the criticism of our work by VSH is groundless.

\end{document}